\documentclass[12pt]{article}

\usepackage{amsmath}
\usepackage{amssymb}
\usepackage{graphicx}
\usepackage{bm}
\usepackage[top=2cm,bottom=2cm,left=2cm,right=2cm]{geometry}

\title{Fast calculation of thermodynamic and structural parameters of solutions using the 3DRISM model and the multi-grid method.
}

\author{
Volodymyr P. Sergiievskyi \\
Max Planck Institute for Mathematics in the Sciences,\\
Inselstrasse 22, DE-04103, Leipzig, Germany\\
e-mail: sergiiev@mis.mpg.de
}

\newcommand{\Angstr}{\AA$~$}

\begin{document}

\maketitle

 { \small \textbf{Note:} the information in this preprint is not up to date. Since the first publication of the preprint (9 Nov 2011) the algorithm was several times modified which allowed to achieve much better computational results.
The paper which describes the new algorithm and new computational tests was recently accepted for the publication in the Journal of Chemical Theory and Computation as: \\
Volodymyr P. Sergiievskyi, Maxim V. Fedorov \\
"3DRISM Multi-grid Algorithm for Fast Solvation Free Energy Calculations" \\
( DOI: 10.1021/ct200815v , http://pubs.acs.org/doi/abs/10.1021/ct200815v )
}

\begin{abstract}

In the paper a new method to solve the tree-dimensional reference interaction site model (3DRISM) integral equations is proposed. The algorithm uses the multi-grid technique which allows to decrease the computational expanses. 3DRISM calculations for aqueous solutions of four compounds (argon, water, methane, methanol) on the different grids are performed in order to determine a dependence of the computational error on the parameters of the grid. It is shown that calculations on the grid with the step 0.05\Angstr and buffer 8\Angstr give the error of solvation free energy calculations less than 0.3 kcal/mol which is comparable to the accuracy of the experimental measurements. The performance of the algorithm is tested. It is shown that the proposed algorithm is in average more than 12 times faster than the standard Picard direct iteration method.

\end{abstract}

\section{Introduction}

Integral equation theory of liquids (IETL) is an effective method for the prediction of the structural and thermodynamic parameters of liquid and amorphous matter \cite{Hansen2000,Martynov1992}.
IETL describes the stucture of the liquid by using the correlation functions.
 The main equation of IETL is the Ornstein-Zernike (OZ) equation which for the case of molecular liquids connects the correlation functions of six independent variables \cite{Hansen2000}.
Because of the comutational complexity solving the OZ equation in a general case is still an open task. In practice one uses approximate models, such as the reference interaction site model (RISM) \cite{Chandler1972} and three-dimensional reference interaction site model (3DRISM) \cite{Beglov1995}.
It was shown that the RISM equation as the implicit solvent model gives more physically-correct description of the system in comparison to the continuum electrostatics models \cite{Ten-no1993,Sato1996,Yokogawa2007}. 
The important application of the RISM equations is the calculation of the solvation free energy (SFE) of compounds.
Within the framework of the proposed model solvation free energy can be analytically calculated from the solutions of the RISM equations \cite{Singer1985,Ten-no2001}.
Recently there were proposed several parameterization methods of the results of the RISM calculations which predict the solvation free energy with the accuracy of 1 kcal/mol \cite{Palmer2010,Ratkova2010,Ratkova2011,Sergiievskyi2011a,Chuev2007c}. 
However, despite the good results RISM still describes not enough accurately the molecular solvent, which leads to the numerous corrections needed for achieving the desired accuracy of the calculations. 
The 3DRISM equations describe more accurately e the stucture of the molecule and, as it was shown, they need essentially less corrections for accurate calculation of the thermodynamic parameters \cite{Palmer2011,Frolov2011a}.
Direct Picard iteration method is the standard algorithm for solving the equations of the IETL \cite{Monson1990}. Although it is easy to implement, but it has quite a low convergence rate.
There were proposed several alternative schemes which increase the convergence rate \cite{Hansen2000,Booth1999,Homeier1995,Kovalenko1999}.
Another way to speed up the calculations is to use the multi-scale methods \cite{Gillan1979,Labik1985,Woelki2008,Kelley2004,Chuev2004,Chuev2004a,Fedorov2005,Fedorov2007a}. 
It is necessary to note that although most of the multi-scale methods use several grids, not all of them do it effictively. However, the multi-grid technology allows to use all the advantages of the multi-scale approach \cite{Hackbusch1985}.
It was recently shown that the multi-grid allows to increase performance of the RISM calculations up to the several dozen times\cite{Sergiievskyi2011}.
In the current paper the multi-grid algorithm which solves the 3DRISM equations is described. The optimal discretization parameters for the solvation free energy calculations are found. The benchmarking of the algorithm and the comparison to the standard Picard iteration method is perfromed. 

\section{Method}
\subsection{3DRISM}

In the paper the 3DRISM for the description of the infinitely diluted solutions is used.
The solvent molecules are given in the RISM approximation, while the solute molecule is a three-dimensional object.
The solvent structure is described by the total and direct correlation functions    of the solvent site $\alpha$: $h_{\alpha}(\mathbf{r})$,  $c_{\alpha}(\mathbf{r})$.
3DRISM equations are written as following:
\begin{equation}
\label{eq:3DRISM}
h_{\alpha}(\mathbf{r}) =
\sum_{\xi=1}^{N_{\rm solvent}}
\int_{\mathbb{R}^3} 
c_{\xi}(\mathbf{r'})\chi_{\xi\alpha}(\mathbf{r'} - \mathbf{r})d\mathbf{r'}
\end{equation}
where $N_{\rm solvent}$ is the number of solvent sites, $\chi_{\xi\alpha}(r)$ is the solvent susceptibility function.
The functions $\chi_{\xi\alpha}(r)$ are defined by the formula $\chi_{\xi\alpha}(r) = \omega_{\xi\alpha}(r) + \rho h_{\xi\alpha}(r)$, where $h_{\xi\alpha}(r)$ is the total correlation function of the pure solvent, $\rho$ is the solvent number density (number of particles in the unit volume), $\omega_{\xi\alpha}(r) = \delta_{\xi\alpha}\delta(r) + (1-\delta_{\xi\alpha})\delta(r-r_{\xi\alpha})(4\pi r_{\xi\alpha}^2)^{-1}$, $\delta_{\xi\alpha}$ is the Kronecer's delta, $\delta(r)$ is the Dirac delta fuunction, $r_{\xi\alpha}$ is the distance between the  sites $\xi$ and $\alpha$ in the solvent molecule.
Equation \eqref{eq:3DRISM} is completed with the closure relation:
\begin{equation}
\label{eq:closure}
h_{\alpha}(\mathbf{r})=
e^{
    -\beta U_{\alpha}(\mathbf{r})
   +h_{\alpha}(\mathbf{r}) - c_{\alpha}(\mathbf{r})
   + B_{\alpha}(\mathbf{r})
}
-1
\end{equation}
Where $\beta=1/k_BT$, $k_B$ is the Boltzmann constant, T is the temperature, $U_{\alpha}(\mathbf{r})$ is the interaction potential of the site $\alpha$ with the solute molecule.
To solve the 3DRISM equations iteratively the equations \eqref{eq:3DRISM} are written in the following form \cite{Perkyns2010}:
\begin{equation}
\label{eq:3DRISM gamma}
\gamma_{\alpha}(\mathbf{r}) =
 \sum_{\xi=1}^{N_{\rm solvent}}
	\int_{\mathbb{R}^3} 
	\mathcal{C}[\gamma_{\xi}(\mathbf{r'}-\mathbf{r})]
	\cdot 
	\chi_{\xi\alpha}(\mathbf{r'})d\mathbf{r'} 
+\theta_{\alpha}(\mathbf{r}) 
- \mathcal{C}[\gamma_{\alpha}(\mathbf{r})]
\end{equation}
where $\gamma_{\alpha}(\mathbf{r}) = h_{\alpha}(\mathbf{r}) – c_{\alpha}^S(\mathbf{r})$,
$c_{\alpha}^S(\mathbf{r}) = c_{\alpha}(\mathbf{r}) + \beta U_{\alpha}^L(\mathbf{r})$,
$U_{\alpha}(\mathbf{r}) = U_{\alpha}^S(\mathbf{r}) + U_{\alpha}^L(\mathbf{r})$,
$U_{\alpha}^S(\mathbf{r})$ is the short-term component of the potential, $U_{\alpha}^L(\mathbf{r})$ is the long-term component of the potential, 
$\theta_{\alpha}(\mathbf{r}) = -\beta \sum_{\xi}\int_{\mathbb{R}^3}U_{\xi}^L(\mathbf{r}-\mathbf{r'})\chi_{\xi\alpha}(\mathbf{r'})d\mathbf{r'}$
In the current paper the Kovalenko-Hirata closure is used \cite{Monson1990}. It is defined by the following formula:
\begin{equation}
\label{eq:KH closure}
\mathcal{C}[\gamma_{\alpha}(\mathbf{r})] = 
\left\lbrace
\begin{array}{l l}
     e^{-\beta U^S_{\alpha}(\mathbf{r}) + \gamma_{\alpha}(\mathbf{r}) } - \gamma_{\alpha}(\mathbf{r}) - 1
    & \text{когда}~~-\beta U^S_{\alpha}(\mathbf{r}) + \gamma_{\alpha}(\mathbf{r})>0 \\
    -\beta U_{\alpha}^S(\mathbf{r}) 
    & \text{иначе}
\end{array}
\right.
\end{equation}

\begin{figure}
\caption{\label{fig:grid} 
The spacing is the distance between the  grid points, and the buffer is the minimal distance from the solute atoms to the grid boundaries
 }
\includegraphics[width=0.5\textwidth]{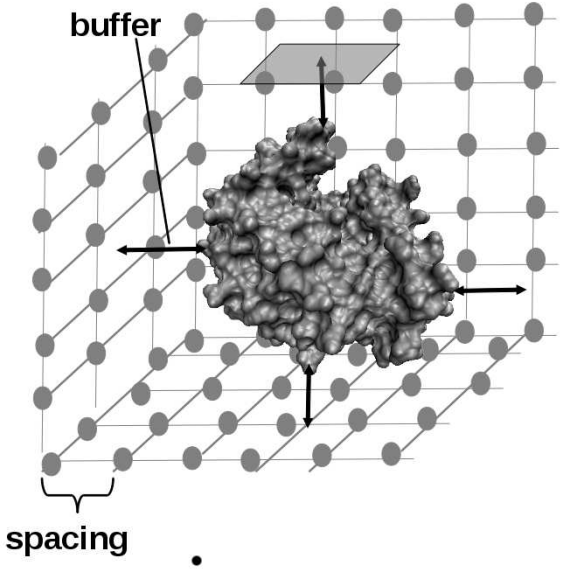}
\end{figure}

In the numerical formulation of the problem \eqref{eq:3DRISM}-\eqref{eq:closure}  the functions $\gamma_{\alpha}(\mathbf{r})$, $\chi_{\xi\alpha}(\mathbf{r})$, $\theta_{\alpha}(\mathbf{r})$ are given by their values in the grid nodes.
We use two values to define the grid:  \emph{spacing} and  \emph{buffer}.
The spacing is the distance between the grid points, and the buffer is the minimal distance from the solute atoms to the grid boundaries (see Figure \ref{fig:grid})

We use the notations $\mathcal{F}_{\mathcal{G}}[\cdot]$, $\mathcal{F}_{\mathcal{F}}^{-1}[\cdot]$ for the forward and inverse discrete Fourier transforms on the grid $\mathcal{G}$.  The discrete equation which corresponds to the equations \eqref{eq:3DRISM gamma} is the following:
\begin{equation}
\label{eq:3DRISM Matrix}
	\bm{\Gamma}^{\mathcal{G}} =
	\mathcal{F}^{-1}_{\mathcal{G}} 
	\left[
		\mathbf{\hat X} \cdot
		\mathcal{F}_{\mathcal{G}}
		\left[
			\mathcal{C}
			\left[
				\bm{\Gamma}^{\mathcal{G}}
			\right]
		\right]	
	\right]
	+ \bm{\Theta}^{\mathcal{G}} 
	- \mathcal{C}
			\left[
				\bm{\Gamma}^{\mathcal{G}}
			\right]
\end{equation}
where $\bm{\Gamma}^{\mathcal{G}} =\left( \bm{\gamma}_1^{\mathcal{G}},\dots,\bm{\gamma}_{N_{\rm solvent}}^{\mathcal{G}}\right)^{T}$,	$\bm{\Theta}^{\mathcal{G}} =	\left(	\bm{\theta}_1^{\mathcal{G}},	\dots,	\bm{\theta}_{N_{\rm solvent}}^{\mathcal{G}} 	\right)^{T}$, $\mathbf{\hat X}^{\mathcal{G}} =  [ \bm{\chi_{\xi\alpha}}^{\mathcal{G}}  ]_{	N_{\rm solvent }\times N_{\rm solvent}	}$, upper index  $\mathcal{G}$ means that the functions are defined by their values in the points of the grid $\mathcal{G}$.
The euqation \eqref{eq:3DRISM Matrix}can be written more compactly:
\begin{equation}
\label{eq:G=FG}
		\bm{\Gamma}^{\mathcal{G}} =
		F[ \bm{\Gamma}^{\mathcal{G}} ]
\end{equation}
where $F[\bm{\Gamma}^{\mathcal{G}}] =
	\mathcal{F}^{-1}_{\mathcal{G}} 
	\left[
		\mathbf{\hat X} \cdot
		\mathcal{F}_{\mathcal{G}}
		\left[
			\mathcal{C}
			\left[
				\bm{\Gamma}^{\mathcal{G}}
			\right]
		\right]	
	\right]
	+ \bm{\Theta}^{\mathcal{G}} 
	- \mathcal{C}
			\left[
				\bm{\Gamma}^{\mathcal{G}}
			\right]$. 
The Picard iteration method is defined by the follwing reccurent formula:
\begin{equation}
\label{eq:Picard}
	\bm{\Gamma}^{\mathcal{G}}_{n+1} =
	(1-\lambda) \bm{\Gamma}^{\mathcal{G}}_{n} +
	\lambda F [\bm{\Gamma}^{\mathcal{G}}_{n}]
\end{equation}
where $\bm{\Gamma}^{\mathcal{G}}_{n}$ - is the $n$-th step solution approximation, $\lambda$ is the mixing parameter. 

\subsection{The multi-grid method}

The multi-grid method is used in the paper to reduce the comutational time \cite{Hackbusch1985,Sergiievskyi2011}.
The problem \eqref{eq:3DRISM gamma} is discretized on several different grids. We say that  the grids which have the smaller number of points the ``coarse'' grids, and the grids which have larger number of points are the ``fine'' grids.
We introduce the operators $p[\cdot]$, $r[\cdot]$ which convert the coarse grid to the finer one and the finer grid to the coarser one correspondingly. We introduce the operator $R[\cdot]$ which transforms the solution $\bm{\Gamma}^{\mathcal{G}}$ from the fine grid $\mathcal{G}$ to the coarse grid $r[\mathcal{G}]$.  
\begin{equation}
	R[\bm{\Gamma}^{\mathcal{G}}] = \bm{\Gamma}^{r[\mathcal{G}]}
\end{equation}
We introduce also the operator $P[\cdot]$, which interpolates the coarse-grid solution $\bm{\Gamma}^{\mathcal{G}}$ to the fine grid $\mathcal{G}$.
\begin{equation}
	P[\bm{\Gamma}^{r[\mathcal{G}]}] = \bm{\Gamma_1}^{\mathcal{G}}
\end{equation} 
In our work the linear interpolation operator $P[\cdot]$ is used.
The multi-grid method solves the following problem:
\begin{equation}
\label{eq:G=FG+D}
	\bm{\Gamma}^{ \mathcal{G}} = F^{\mathcal{G}}[\bm{\Gamma}^{\mathcal{G}}] + \mathbf{D}^{\mathcal{G}}
\end{equation}
where $\mathbf{D}=\left( \mathbf{d}^{\mathcal{G}}_{1},\dots, \mathbf{d}^{\mathcal{G}}_{N_{\rm solvent}}\right)^{T}$ is the vector of corrections.
	
It is convenient to introduce the iterative operator $\Lambda_{\mathcal{G}}[\cdot;\cdot]$, which is defined in the following way:
\begin{equation}
	\Lambda_{\mathcal{G}}[\bm{\Gamma}_{\mathcal{G}};\mathbf{D}^{\mathcal{G}}] =
	(1-\lambda)\bm{\Gamma}^{\mathcal{G}} +
	\lambda \left( F_{\mathcal{G}}[\bm{\Gamma}^{\mathcal{G}}] + \mathbf{D}^{\mathcal{G}} \right)
\end{equation}

The iterative multi-grid algorithm which solves the equation \eqref{eq:G=FG+D} can be written in a following way:
\begin{equation}
	\bm{ \Gamma}_{n+1}^{\mathcal{G}} = 
	\mathcal{M}_{\mathcal{G}}^l
	\left[
		\bm{\Gamma}_{n}^{\mathcal{G}};
		\mathbf{D}^{\mathcal{G}}
	\right]
\end{equation}
where $\bm{\Gamma}_{n}^{\mathcal{G}}$ is the solution approximation on the $n$-th step, $\mathcal{M}_{\mathcal{G}}^l [\cdot;\cdot]$ is the multi-grid operator, which performs one multi-grid iteration step of depth $l$ on the grid $\mathcal{G}$.
Calculation of the multi-grid operator of depth $l=0$ is equal  to the performing of $n$ one-grid iteration steps on the grid $\mathcal{G}$
If  $l>0$ for the given initial guess $\bm{\Gamma}_n^{\mathcal{G}}$ and correctons vector $\mathbf{D}^{\mathcal{G}}$ the multi-grid operator $\mathcal{M}_{\mathcal{G}}^l [\cdot;\cdot]$ is calculated by the following algorithm:

\textbf{Input}: $\bm{\Gamma}_{n}^{\mathcal{G}}$, $\mathbf{D}^{\mathcal{G}}$, $l$

\textbf{Output}: $\bm{\Gamma}_{n+1}^{\mathcal{G}} = \mathcal{M}_{\mathcal{G}}^l [ \bm{\Gamma}_{n}^{\mathcal{G}}; \mathbf{D}^{\mathcal{G}} ]$

\begin{enumerate}

\item Perform $\nu_1$ Picard iteration steps on the fine grid (in our work $\nu_1=1$):
\begin{equation*}
	\bm{\Gamma'}^{\mathcal{G}} = \left( \Lambda_{\mathcal{G}} \right)^{\nu_1}
	\left[
		\bm{\Gamma}_{n}^{\mathcal{G}};
		\mathbf{D}^{\mathcal{G}}
	\right]
\end{equation*}

\item Move to the coarse grid $r[\mathcal{G}]$:
\begin{equation*}
	\bm{\Gamma}^{r[\mathcal{G}]}_{(0)} = R[\bm{\Gamma'}^{\mathcal{G}}] 
\end{equation*}

\item Calculate the coarse grid correction: 
\begin{equation*}
	\mathbf{E}^{r[\mathcal{G}]} = 
	R \left[
		F[\bm{\Gamma'}^{\mathcal{G}}]
	\right]
	-
	F[\bm{\Gamma}_{(0)}^{r[\mathcal{G}]}]
\end{equation*}

\item Repeat recoursively $\mu$ multi-grid iteration steps of depth $l-1$ on the coarse grid (in our work  $\mu$=1):
\begin{equation*}
	\bm{\Gamma}^{r[\mathcal{G}]}_{(\mu)} = 
	\left(
		\mathcal{M}_{r[\mathcal{G}]}^{l-1}
	\right)^{\mu}
	\left[
		\bm{\Gamma}^{r[\mathcal{G}]}_{(0)};
		R[\mathbf{D}^{\mathcal{G}} ]+ 
		\mathbf{E}^{r[\mathcal{G}]}		
	\right]
\end{equation*}

\item Correct the fine-grid solution:
\begin{equation*}
	\bm{\Gamma''}^{\mathcal{G}} = 
	\bm{\Gamma'}^{\mathcal{G}}
	+ P 	\left[
			\bm{\Gamma}^{r[\mathcal{G}]}_{(\mu)} -
			\bm{\Gamma}^{r[\mathcal{G}]}_{(0)}	
		\right]
\end{equation*}

\item Perform $\nu_2$ Picard iteration steps on the fine grid (in our work $\nu_2=0$):
\begin{equation*}
	\bm{\Gamma}^{\mathcal{G}}_{n+1} =
	\left(	\Lambda_{\mathcal{G}}
	\right)^{\nu_2}
	\left[
		\bm{\Gamma''}^{\mathcal{G}};
		\mathbf{D}^{\mathcal{G}}
	\right]
\end{equation*}	

\end{enumerate}

The iterations are performed until the following stop-condition satisfies:
\begin{equation}
	|| \bm{\Gamma}_{n} - \bm{\Gamma}_{n+m} || < \varepsilon_{\rm tres}
\end{equation}
where $m$ is selected in such a way that the following relation holds:
\begin{equation}
	||\bm{\Gamma}_{n+m}^{\mathcal{G}} - \bm{\Gamma}_{n+m+1}^{\mathcal{G}}|| < 
	0.01 ||\bm{\Gamma}_{n}^{\mathcal{G}} - \bm{\Gamma}_{n+1}^{\mathcal{G}} ||
\end{equation} 

In the paper is used the norm based on the solvation free energy calculations:
\begin{equation}
	|| \bm{\Gamma}^{\mathcal{G}}_1 - \bm{\Gamma}^{\mathcal{G}}_2|| =
	| \Delta G_{KH}(\bm{\Gamma}_1) - \Delta G_{KH}(\bm{\Gamma}_2) |
\end{equation}
where the solvation free energy is calculated using the Kovalenko-Hirata formula \cite{Hirata2003}:
\begin{equation}
\label{eq:DeltaG}
	\Delta G_{KH}(\bm{\Gamma}^{\mathcal{G}}) =
	\rho k_B T \sum_{\alpha}^{N_{\rm solvent} } 
	\int_{\mathbb{R}^3}
	\theta(-h_{\alpha}(\mathbf{r}))	h_{\alpha} (\mathbf{r}) -
	\frac{1}{2}c_{\alpha}(\mathbf{r})h_{\alpha}(\mathbf{r}) -
	c_{\alpha}(\mathbf{r})
	d\mathbf{r}
\end{equation}
In the paper the value theshold value $\varepsilon_{\rm tres}$=0.001 is used.

\subsection{Computational details}

For the argon, methane, metanol and water the OPLS-AA partial charges were used\cite{Jorgensen1996a}.
For the water (solute and solvent species) the parameters of the MSPC-E water model were used \cite{Fedorov2007}.
In the paper the total correlation functions of the pure water calculated in the paper \cite{Fedorov2007} were used.
Pairwise $\sigma$ Lennard-Jones parameters were calculated as the arithmetic mean of the atomic parameters, pairwise $\varepsilon$ Lennard-Jones parameters were calculated as the geometric mean of the atomic parameters:
\begin{equation}
\sigma_{12}=\frac{\sigma_1 + \sigma_2}{2} \qquad 
\varepsilon_{12} = \sqrt{\varepsilon_1\cdot\varepsilon_2}
\end{equation}
Calculations were performed on the Dual Core AMD Opteron(tm) Processor 885 with the clocking  2613 MHz. The forward and inverse Fourier transforms were calculated using the FFTW3 library \cite{Frigo1999}.

\section{Results}

\begin{figure}[h!]
\caption{\label{fig:dr} Dependence of the error of the SFE calculations  (Error of ΔG) on the grid spacing  }
\includegraphics[width=0.5\textwidth]{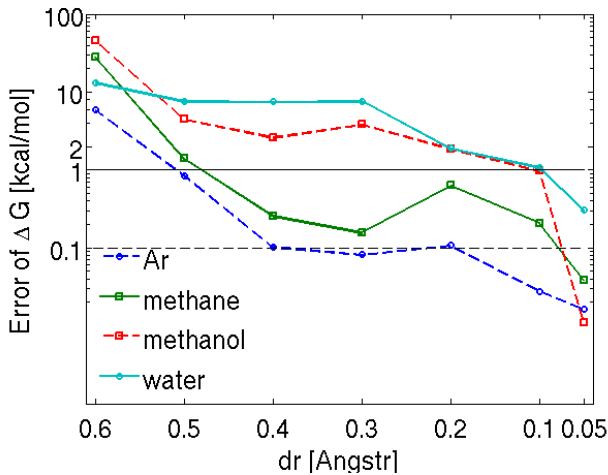}
\end{figure}

The 3DRISM calculations were performed for the infinitely dilutes aqueous solutions of four compounds: argon, methane, methanol, and water.
To determine the optimal grid parameter the solvation free energy calculations for the grids with the different spacing and buffer were performed.
In the Figure \ref{fig:dr} the dependence of the calculation error on the grid spacing is presented.
For the calculations the grids with the buffer 5\Angstr and the spacing from 0.05\Angstr to 0.6\Angstr were used.
The calculation error on each grid was calculated as the absolute value of the difference between the SFE calculated on the current grid and the SFE calculated on the  grid with the spacing 0.025\Angstr. The solvation free energy values calculated on this grid were used as the standard. 
On the grid with the spacing 0.05\Angstr the maximal calculation's error is less than 0.3 kcal/mol, while for the grids with the larger number of points the calculation's error is more than 1 kcal/mol.
Relatively large calculations' errors can be explained by the fact that the Cartesian grid cannot approximate the spherically symmetrix functions $\omega_{\xi\alpha}(r)$ which define the structure of the solvent molecule with the good accuracy.
Thus in a fact we use the solvent molecules with the distorted geometry which causes the decreasing of the  accuracy of the calculations.
The acceptable accuracy for the chemical applications is 0.5 kcal/mol.
As it was shown, it is enough to have the grid with the spacing 0.05\Angstr in order to calculate the SFE with the accuracy 0.5 kcal/mol.

\begin{figure}
\caption{\label{fig:buffer}  Dependence of the error of the SFE calculations (Error of ΔG) on the grid buffer.}
\includegraphics[width=0.5\textwidth]{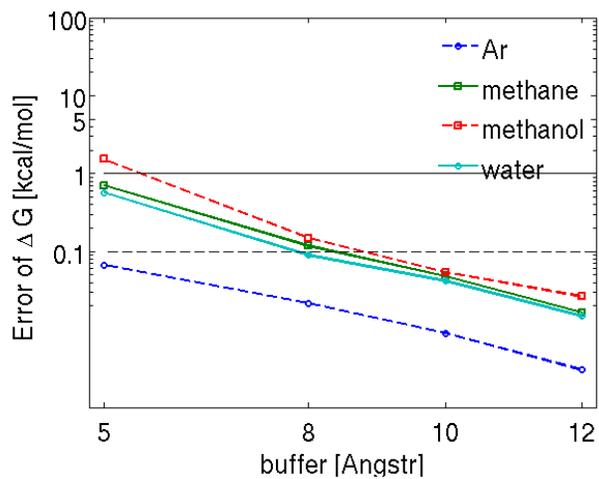}
\end{figure}

In  Figure \ref{fig:buffer} the dependence of the calculation error on the grid buffer is presented. Calculations were performed on the grids with the constant spacing $\Delta R$=0.05\Angstr.
The error was calculated as the difference between the SFE calculated on the current grid and the SFE calculated in the grid with the spacing 0.05\Angstr and the buffer 14\Angstr which was used as the standard.
For the calculations with the accuracy of 0.5 kcal/mol it is enough to use the buffer 8\Angstr.

To test the performance of the proposed algorithm  the calculations for the aqueous solutions of argon, methane, methanol, and water on the grid with the spacing 0.05\Angstr and the buffer 8\Angstr were performed.
The multi-grid algorithm with the depth $l=2$ was used.
In Table 1 the  comparison of the comutation time of the multi-grid and Picard iteration methods is presented.
The average computation time for the Picard iteration method is 7 h. 42 min., while the average computation time for the proposed multi-grid method is 36 min. In average the multi-grid method is 12.2 times faster than the Picard iteration method.

\begin{table}
\caption{\label{tab:time}
Comparison of the comutation time for the  Picard iteration method and for the multi-grid method.
 }

\begin{tabular} { c | c  | c |c }
\hline 
Compound & Picard iteration & multi-grid & speedup (times) \\ 
\hline
argon & 4 h. 45 min.   & 25 min. & 11.5  \\
methane & 10 h. 42 min.   & 41 min. & 15.7  \\
methanol & 12 h. 16 min. & 48 min. & 15.3 \\
water & 3 h. 5 min.      & 29 min. & 6.4 \\
\\  
\hline
\textbf{Average} & 7 h. 42 min. & 36 min. & 12.2
\end{tabular}
\end{table}

\section{Conclusions}

In the paper the new multi-grid based method for solving the 3DRISM equations is proposed.
To determine the optimal grid parameters the 3DRISM calculations for infinitely diluted aqueous solution of argon, methane, methanol and water were performed.
It is shown that on the gird with the spacing 0.05\Angstr and the buffer 8\Angstr the maximal comutational error is less than 0.3 kcal/mol, while for the grids with the larger spacing the computational error is more than 1 kcal/mol.
The performance of the proposed algorithm was compared to the performance of the standard Picard iteration method.
It is shown that the average calculation time is 36 min. for the proposed method while the average calculation time for the Picard iteration method is more than 7 hours.
Thus, we show that the multi-grid algorithm is more than 12 times faster than  the standard Picard method.

\section{Acknowledgements}

I would like to acknowledge my supervisor, Maxim V. Fedorov, for the organizing the research. 
I would like to acknowledge Olena E. Prokof'eva for reading the manuscript and useful comments.
I also would like to acknowledge the Max Planck Institute for Mathematics in the Sciences for the financial support of research.


\end{document}